\documentclass[conference]{IEEEtran}
\IEEEoverridecommandlockouts
\usepackage{cite}
\usepackage{url}
\usepackage{amsmath,amssymb,amsfonts}
\usepackage{algpseudocode}
\usepackage{graphicx}
\usepackage{textcomp}
\usepackage{xcolor}
\usepackage{enumitem}
\usepackage{multirow}
\usepackage{array}
\usepackage{ctable}
\def\BibTeX{{\rm B\kern-.05em{\sc i\kern-.025em b}\kern-.08em
    T\kern-.1667em\lower.7ex\hbox{E}\kern-.125emX}}
\begin{document}

\algnewcommand{\Initialize}[1]{%
  \State \textbf{initialize:}
  \Statex \hspace*{\algorithmicindent}\parbox[t]{.8\linewidth}{\raggedright #1}
}
\algnewcommand{\Input}[1]{%
  \State \textbf{input:}
  \Statex \hspace*{\algorithmicindent}\parbox[t]{.8\linewidth}{\raggedright #1}
}
\algnewcommand{\Output}[1]{%
  \State \textbf{output:}
  \Statex \hspace*{\algorithmicindent}\parbox[t]{.8\linewidth}{\raggedright #1}
}
\title{Accelerated carrier invoice factoring using predictive freight transport events\\
}

\makeatletter
\newcommand{\linebreakand}{%
  \end{@IEEEauthorhalign}
  \hfill\mbox{}\par
  \mbox{}\hfill\begin{@IEEEauthorhalign}
}
\makeatother

\author{
\IEEEauthorblockN{Krishnasuri Narayanam}
\IEEEauthorblockA{\textit{IBM Research, India} \\
knaraya3@in.ibm.com}
\and
\IEEEauthorblockN{Pankaj Dayama}
\IEEEauthorblockA{\textit{IBM Research, India} \\
pankajdayama@in.ibm.com}
\and
\IEEEauthorblockN{Sandeep Nishad}
\IEEEauthorblockA{\textit{IBM Research, India} \\
sandeep.nishad1@ibm.com}
}


\maketitle

\IEEEpubidadjcol

\begin{abstract}
Invoice factoring is an invoice financing process where business organizations sell their invoices to banks or financial institutions at a discount to gain faster access to the invoice amount. In global trade, ocean and land carriers exercise invoice factoring to gain quick access to the money they get paid for the shipment of consignments by shippers. Shippers typically clear the invoice payment within 60-90 days of goods getting delivered. In order to get early access to capital, carriers initiate invoice factoring after the completion of goods delivery to the shipper. In this work, we provide an approach to enable accelerated carrier invoice factoring even before the delivery of goods. Carrier invoice value for a given shipment depends on the actual values of shipment tracking events. We predict the carrier invoice value at different freight transport milestone events as the goods transportation progresses from supplier to shipper using smart contracts on a blockchain network that is operated by the global trade logistics participants. The prediction at a given stage is based on past shipment tracking events and predicted future shipment tracking events. Accurate prediction of invoice value for ongoing shipment enables the carrier organization to initiate invoice factoring on the trade finance network before the completion of goods delivery to the shipper. Further, based on the past accuracy of prediction models, the financial institutions may choose to release the invoice amount in installments at different freight transportation milestone events.
\end{abstract}

\begin{IEEEkeywords}
Global trade, event prediction, invoice computation, invoice factoring, smart contract, blockchain interoperability.
\end{IEEEkeywords}

\section{Introduction}
\label{sec:introduction}

Global trade is a supply chain transaction in which shippers buy goods from suppliers with goods transportation across international boundaries provided by the carriers. A single trade transaction involves the execution of processes like freight transportation, invoice generation, receiving of goods, invoice processing, and payment processing \cite{NGSSSICBC21}. One or more carrier organizations may be responsible for the freight transport (e.g., origin land carrier, ocean carrier, destination land carrier) across country boundaries in global trade. Shippers are issued invoices from suppliers for the goods provided and from carrier organizations for the goods transportation. The accounts payable team of the shipper verifies the invoices and carries out payment processing to the supplier and all the carriers.

Payment processing involves executing a payment method as per the terms captured in the service contracts between trade participants (i.e., shippers, suppliers, and carriers). One of the payment methods is {\em invoice financing} that is facilitated by banks or financial institutions. {\em Invoice factoring} \cite{factoring} is used to finance the carrier invoices and {\em letter of credit} \cite{letterOfCredit} is used to finance the supplier invoices in global trade.

An alternate payment method is {\em open account} \cite{openaccount} where payment processing takes place typically in 30, 60, or 90 days after the completion of the goods delivery by the carrier. Carrier organizations obtain cash soon after goods delivery completion using invoice factoring against the many days of delay in the open account method.

With invoice factoring, the factoring organization pays the invoice amount with a discount to the carrier, typically within a day post goods shipment and delivery completion to the shipper. The factoring company collects its fee as a discount in the invoice amount. The shipper pays the invoice amount to the factoring company, typically within 60 days after goods delivery. This benefits each party involved in as follows:
\begin{itemize}[leftmargin=*]
    \item Carrier: gains quicker access to money it owed soon after goods delivery
    \item Shipper: can get more time from the factoring company to pay off the invoice amount
    \item Factoring organization: earns by invoice discount
\end{itemize}

In this paper, we propose accelerating the carrier invoice factoring such that the carrier organizations obtain cash even before goods delivery completion to the destination. We achieve this by predicting the carrier invoices at different freight transport milestones as the goods transportation progresses from the origin to the destination and initiating the invoice factoring based on the predicted invoices before goods delivery completion. Our proposal further increases the benefits of invoice factoring for both the carrier and factoring organizations, while the shipper doesn't experience any changes from the existing invoice factoring behavior. In particular, carriers get access to the invoice amount soon after the goods dispatch and much before goods delivery to the destination, and factoring organizations can get higher invoice discounts.

\subsection{Illustrative example}
We illustrate the advantages to the carriers and the factoring organizations due to the payment processing method that we propose, namely {\em accelerated invoice factoring}, using a simple example that considers various payment processing methods. 
\begin{figure}[ht]
    \centering
    \includegraphics[width=1.0\columnwidth]{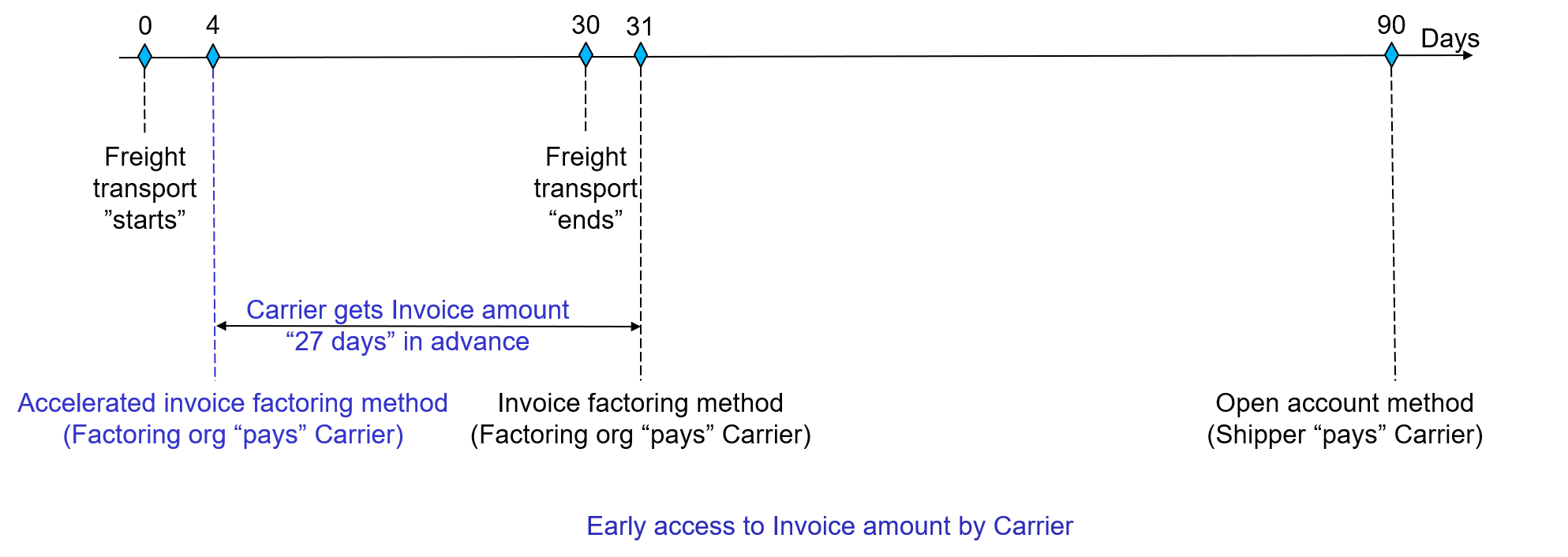}
    \caption{Payment processing methods}
    \label{fig:paymentmethods}
\end{figure}

Assume that the goods transportation from the origin to the destination takes $30$ days for the carrier and the invoice amount is $100$ USD. Below are sample payment settlement details under different payment processing methods.
\begin{enumerate}[label=\roman*,leftmargin=\parindent]
    \item {\em Open Account method}: 
    Carrier receives the invoice amount of $\$100$ directly from the shipper, after $60$ days of the goods delivery to the destination (i.e., on $90^{th}$ day after goods dispatch by the supplier).
    \item {\em Invoice Factoring method}: Carrier receives the invoice amount after $1$ day of goods delivery to the destination (i.e., on $31^{st}$ day of goods dispatch). However, the carrier only receives a discounted invoice amount of $\$94$ from the factoring organization. The shipper pays the invoice amount of $\$100$ to the factoring organization after $60$ days of the goods delivery (i.e., on $90^{th}$ day after goods dispatch by the supplier).
    \item {\em Accelerated Invoice Factoring method}: Carrier receives the invoice amount after $4$ days of goods dispatch by the supplier and before the completion of goods transportation to the destination. However, the carrier only receives a discounted invoice amount of $\$92$ from the factoring organization. The shipper pays the invoice amount of $\$100$ to the factoring organization after $60$ days of good delivery (i.e., on $90^{th}$ day after goods dispatch by the supplier).
\end{enumerate}
Please note that the invoice discount amount is more in our proposed method. The factoring organization gains an additional $\$2$ (i.e., the difference between $\$94$ as per invoice factoring method minus $\$92$ as per our proposed method that it pays to the carrier).  The trade-off here is that the factoring organization needs to release the discounted invoice amount $27$ days early (i.e., on $31^{st}$ day after goods dispatch as per invoice factoring method minus on $4^{th}$ day after goods dispatch as per our proposed method).

We organize the rest of the paper as follows. In Section-\ref{sec:relatedWork}, we provide the related work and references to blockchain-based solutions related to supply chain optimization that helps realize our contributions. In Section-\ref{sec:background}, we present details about blockchain-based carrier invoice generation and information on carrier invoice factoring. In Section-\ref{sec:design}, we describe the details of our proposed blockchain-based payment settlement method. In Section-\ref{sec:implementation}, we discuss an approach to realize the proposed {\em accelerated invoice factoring} method using state-of-art blockchain-based supply chain networks. Section \ref{sec:modeltraining} presents the details on how we have implemented an AI model for event occurrence prediction and some guidelines on the same. Section \ref{sec:conclusion} summarizes our contributions.

\section{Related Work}
\label{sec:relatedWork}
Study on the use of blockchain for supply chain optimization is an evolving research domain. There exist blockchain-based trade logistic networks like TradeLens \cite{TLIBM} that help in real-time shipment tracking. There exist blockchain-based platforms that generate carrier e-invoices in real-time using the shipment tracking details and service contract agreements \cite{NGSSCSCVBlockchain20}. Similarly, there exist blockchain-based accounts payable systems \cite{NGSSSICBC21,bcforap,bcinar} that help in invoice processing. And there are blockchain-based trade finance systems \cite{wt,marcopolo,chang2020blockchain,chiu2019blockchain,bogucharskov2018adoption} that help in payment processing to the suppliers and carriers. Not all trade finance systems use the same payment processing method, and there are many different types of them. All these solutions use blockchain technology to address various process inefficiencies at different stages of the global trade supply chain process. We propose to leverage this suite of blockchain-based systems to enhance the gains of the supply chain participants via our payment processing method called {\it accelerated invoice factoring}.
\section{Background}
\label{sec:background}

In this section, we provide a summary of auto-generation of blockchain-based carrier invoices \cite{NGSSCSCVBlockchain20} and carrier invoice factoring in global trade. 
\subsection{Carrier e-invoice generation}
\label{sec:auto-e-invocing}
A carrier invoice is a set of charges involved in the goods transportation process. The shipper and the carrier agree on various freight transport charges that can be part of the carrier invoice and the rate at which these charges add to the invoice. It's called the {\em service contract agreement} and stored on a blockchain-based trade logistics network in which both shipper and carrier are the network peers.

A charge can be planned (applicable to all the shipments) or unplanned (applied if the required container tracking events occur). Moreover, a charge can be fixed (computed directly using a fixed service contract rate) or variable (computed using the service contract rates and the container tracking events). For example, the demurrage charge \cite{DemurrageCharge} is an unplanned variable charge component of an ocean carrier invoice that uses container tracking events to determine the waiting time of the container in a port terminal.

Packed goods are loaded into a container and the carrier ships the container from the origin to the destination. Container tracking events (also known as shipping milestone events) track the progress of the freight transport. These shipping milestone events are also recorded on the trade logistics network, such that it provides visibility into the current state of the container movement to all the trade participants.

Smart contracts on blockchain generate the carrier invoices (also known as e-invoices) using the real-time shipping milestone events and the service contract agreements available on the blockchain ledger. Smart contracts compute the carrier e-invoices iteratively on the occurrence of various shipment milestone events. An e-invoice may be partial and may not contain all the charges until the last milestone event. Examples of shipment milestone events of an ocean carrier are {\em container loaded on the vessel, vessel departure at origin port, vessel arrival at the destination port, or container discharge from vessel}. The ocean carrier invoices include both export port-specific charges and import port-specific charges.

\subsection{Carrier invoice factoring}
\label{sec:carrierinvoicefactoring}
Carrier invoice factoring allows freight transportation organizations (e.g., trucks or ocean carriers \cite{oceanfreightfactoring1, oceanfreightfactoring2, oceanfreightfactoring3}) to receive cash from unpaid invoices immediately. Carrier organizations need to complete the freight delivery and submit the invoice (sometimes along with the bill of lading) to the bank/factoring company. Bank verifies the invoices and releases the invoice amount minus factoring fee (also known as invoice discount amount) within $24$ hours. Invoice factoring speeds up the operating cash flow for carriers that helps grow their business. Without invoice factoring, the carrier organizations need to wait up to $90$ days for the shippers to settle the full invoice amount (without invoice discount). Note that the bill of lading (BOL) is a document that consists of details about the parties involved in a trade transaction. 
Carriers use BOL to generate the invoice and issue it to the shipper.
\section{System design}
\label{sec:design}
This section describes the design of our blockchain system that supports accelerated invoice factoring.

\subsection{Predicted carrier e-invoice generation}
\label{predicted-e-invoices}
In the carrier e-invoice computation, as described in section \ref{sec:auto-e-invocing}, only the actual shipment milestone events are used and the future shipping milestone events are not considered. However, we observe that the estimates of future shipping milestone events might be available from different trade participants, or it may be possible to estimate future milestone events afresh. Thus, we can compute the carrier e-invoices using a combination of the actual values of already occurred milestone events and the predicted values of milestone events that are yet to occur. We call these carrier e-invoices as predicted e-invoices.

Please note that these predicted carrier e-invoices may be partial and may not contain all the charges until the last milestone event occurs, as it may not be possible to predict all the future milestone events. For those future events for which prediction of their occurrence time is possible, the prediction activity can be carried out anytime before the actual occurrence time of the event, say on the actual occurrence of a prior shipment milestone event as the freight transport progresses. Prediction in the current iteration for a milestone event may use all the previously predicted occurrence time values for that milestone event.

We propose to use an AI model to find the aggregate estimated occurrence of a container tracking event. For example, using all the container tracking events till the current milestone event ``{\em vessel departure at the origin port}", the AI model can estimate the immediate future milestone event ``{\em vessel arrival at the destination port}". The AI model also takes care of the inconsistencies between the estimated values of the milestone events if estimates are available from more than one supply chain participant.

Please note that the same future milestone event, {\em vessel arrival at the destination port}, would have got predicted previously by the AI model for this same shipment on the occurrence of the past milestone event {\em container loaded on the vessel}. Essentially the estimation of a shipment tracking event by the AI model can happen multiple times till the actual occurrence of that event. Each time there is a change in the estimated occurrence of a container event, the blockchain smart contract can update the carrier e-invoice by considering all the actual past milestone events, the latest estimated future milestone events, and the service contract agreements.

\subsection{Factoring the predicted carrier e-invoices}
\label{factoring-predicted-e-invoices}
The first predicted carrier e-invoice is available for the goods shipped (consignment) soon after the dispatch of the goods by the supplier. It may be a partial e-invoice and only contain actual values of planned-fixed charges and predicted values for invoice charges of type planned-variable, unplanned-fixed, and unplanned-variable (refer to \ref{sec:auto-e-invocing}).

In our proposed solution, the carrier can present the predicted e-invoice to the factoring organization immediately once it is generated. The factoring organization releases part of the e-invoice predicted amount (first chunk of the invoice amount after discount) soon to the carrier. Whenever the e-invoice predicted amount is updated on the occurrence of a subsequent shipping milestone event as the freight transportation progresses, it is presented again to the factoring organization by the carrier such that an additional chunk of the carrier invoice amount might be released. This process iterates till the occurrence of the last shipping milestone event. After the goods are delivered to the shipper, the blockchain smart contract generates the actual carrier invoice. Carrier presents the actual invoice to the factoring organization for the last time, and the factoring organization releases the final chunk of the discounted invoice amount (if any) within $24$ hours. This factoring process is different from the traditional invoice factoring, where the factoring organization releases the discounted invoice amount in one chunk when the last shipping milestone event occurs.

Please note that if the accuracy of the model used for predicting e-invoice is good, then the factoring organization might choose bigger chunks of the discounted invoice amount in the initial iterations itself.

The computation of the predicted e-invoice happens on the trade logistics network, and the factoring organization carries out payment settlement to the carrier on the trade finance network. Supply chain participants like suppliers, carrier organizations, and shippers are part of the trade logistics network and  financial organizations and carriers are part of the trade finance network.

\subsection{AI model overview}
\label{ai-model-entity}
The AI model used to predict the occurrence time of future milestone events can be an independent service provided by third-party outside the trade logistics network or it can be an independent party of the trade logistics network that provides predicted future milestone events for all (or part of) the shipments tracked on the logistics network. Carrier organizations can pay to the party providing the AI prediction service since the carrier benefits from the proposed {\em accelerated invoice factoring} payment processing method.

Please note that the accuracy of the AI model prediction may vary from one global trade shipping lane to other lane. The accuracy of a milestone event prediction depends on the quality of the input data used by the model.  The input data includes external factors like the capacity of shipping ports (number of terminals at a port), the number of transshipment ports on the shipping lane, traffic at different shipping ports on the shipping lane, sea tide, weather conditions, vessel type, etc. Section \ref{sec:modeltraining} captures more details on the AI model training.


\subsection{Workflow of Accelerated Invoice Factoring}
\label{workflow-invoice-factoring}
Here we describe the invoice financing process workflow as per our proposed {\em accelerated invoice factoring} method.
\begin{figure}[ht]
    \centering
    \includegraphics[width=0.99\columnwidth]{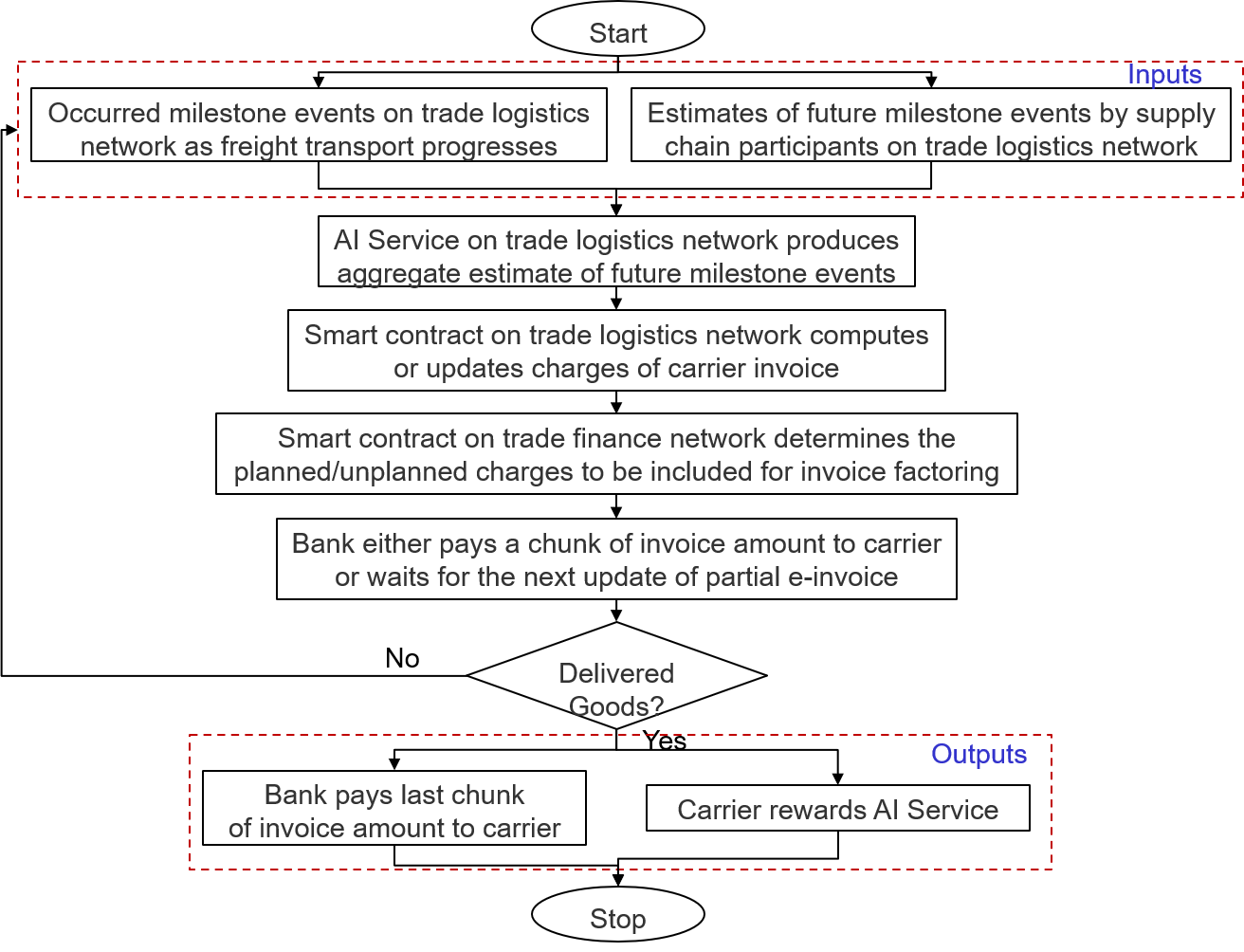}
    \caption{Invoice finance workflow as per accelerated invoice factoring}
    \label{fig:workflowourpaymentmethod}
\end{figure}
 In Figure \ref{fig:workflowourpaymentmethod}, we illustrate the different steps involved in our approach. Below are the details of each step:
\begin{itemize}[leftmargin=*]
    \item The workflow starts when the shipper places a purchase order of goods that the supplier dispatches.
    \item The trade logistics blockchain network tracks the occurred shipping milestone events as the carrier transports the goods from the origin to the destination. Trade logistics blockchain platform also tracks the estimates on future milestone events submitted by various supply chain participants. These estimates on future milestone events help the downstream supply chain participants plan their activities accordingly.
    \item AI service produces an aggregate estimate (refer Section \ref{sec:aggregate-estimate-generation}) of future milestone events using the inputs described in the previous step. 
    \item Smart contract on the trade logistics blockchain platform computes the partial carrier e-invoice for the shipment of the goods using the milestone events (actual \& predicted) and the service contract agreements previously loaded on the blockchain ledger.
    \item Smart contract on the trade finance blockchain platform determines the planned/unplanned charges to be included in the current iteration of the accelerated invoice factoring (refer Section \ref{sec:decision-on-charges-for-factoring}). For the current iteration of the invoice financing, banks consider the actual charges not paid in the previous iterations; and the predicted charges not settled in the iterations until then and that have high prediction accuracy. If the prediction accuracy of an invoice charge is less, then the banks either wait for a better prediction of the charge amount or wait until the addition of the actual charge value to the partial e-invoice.
    \item Bank initiates payment of one more invoice amount chunks to carrier organization if invoice financing in the current iteration includes one or more charges (actual or predicted).
    \item The global trade participants iterate over the above steps at the occurrence of each milestone event until the goods are delivered.
    \item After the goods delivery completion, the bank checks if the sum of the chunks of partial invoice amounts paid across all the iterations is less than the actual invoice amount after discount.  If so, the bank computes the remaining amount to be paid and releases the final chunk of the invoice amount to the carrier organization. If the bank has already transferred more than it's entitled to pay, it adjusts the surplus paid against the future shipments from the same carrier.
    \item After the goods are delivered, the carrier rewards the AI service provider since the carrier organization now got early access to the invoice amount by following our proposed approach.
\end{itemize}


\subsection{AI service for aggregate estimate computation}
\label{sec:aggregate-estimate-generation}

\begin{figure}[ht]
    \centering
    \includegraphics[width=0.99\columnwidth]{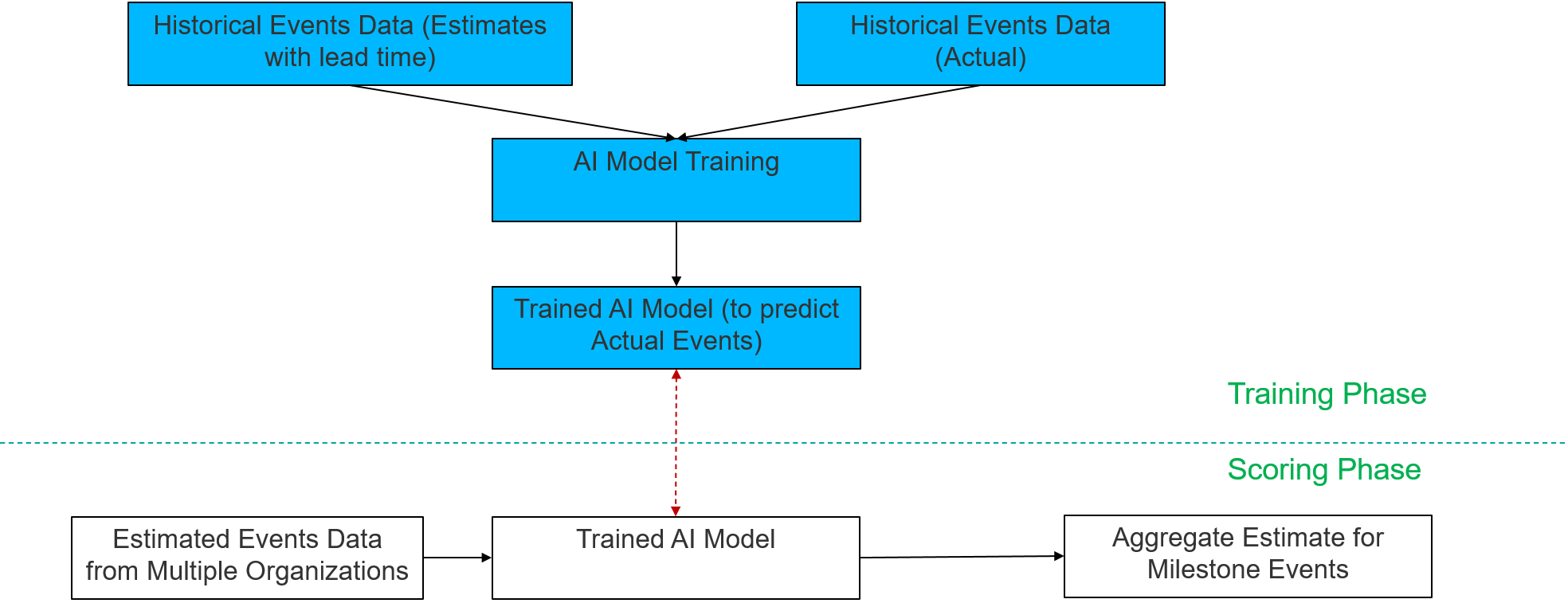}
    \caption{Approach for aggregate estimate computation by AI service}
    \label{fig:aggregate-estimate-generation}
\end{figure}

Here we describe the approach to generate aggregate event occurrence time estimate by the AI service for any milestone event. Note that there could be zero or more estimates from different sources for a given future milestone event (i.e., an event that has not yet happened). Training of the AI model should ideally take place specifically for a given global shipping lane to achieve better prediction accuracy. Figure \ref{fig:aggregate-estimate-generation} captures the steps involved in the proposed approach. Training data includes the historical estimated and actual events data. Section \ref{sec:modeltraining} captures more details on the model training. Note that the training happens outside of the trade logistics blockchain platform.  Testing and scoring of AI model happens on the trade logistics blockchain platform.

\subsection{Decision on charges for factoring consideration}
\label{sec:decision-on-charges-for-factoring}

\begin{figure}[ht]
    \centering
    \includegraphics[width=0.99\columnwidth]{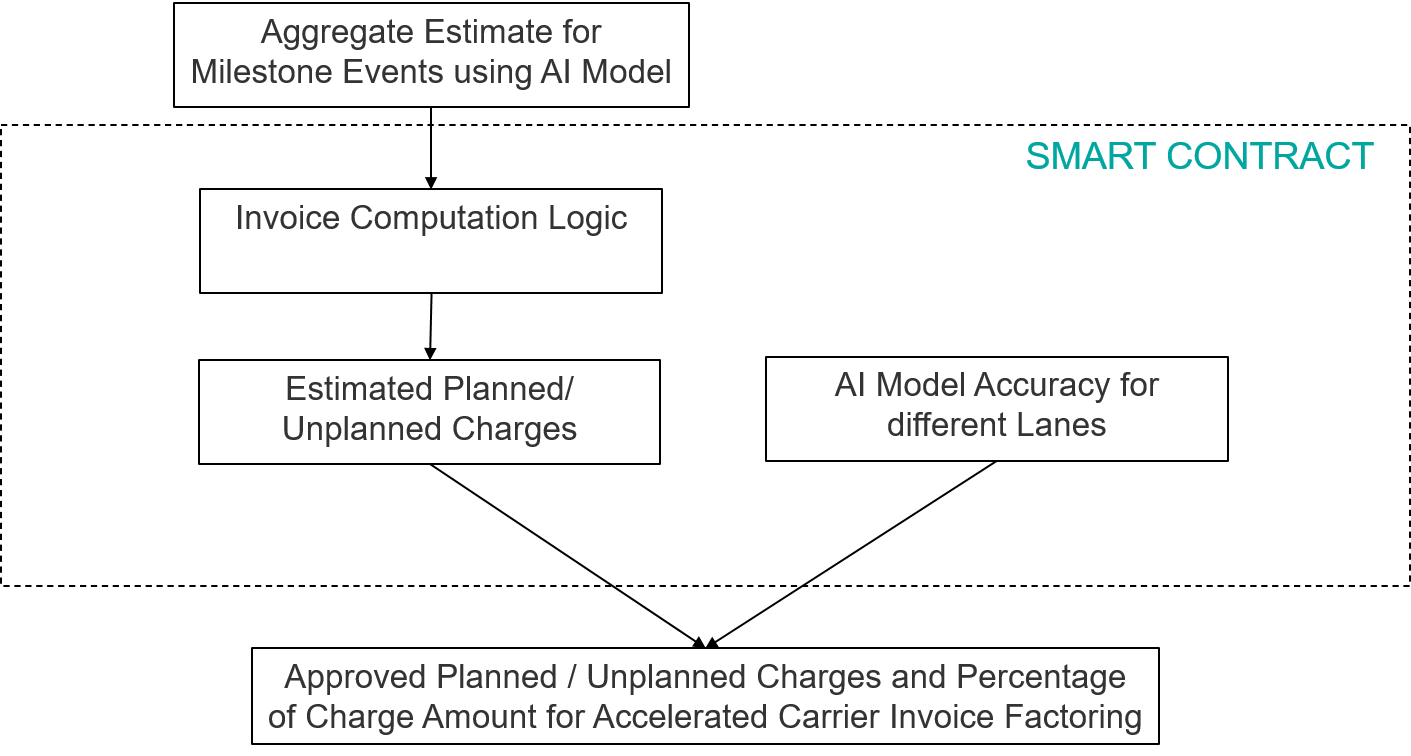}
    \caption{Approach for charge inclusion for accelerated invoice factoring}
    \label{fig:decision-on-charges-for-factoring}
\end{figure}

We describe here the approach to decide on the charges considered for {\em accelerated invoice factoring}. Note that a charge included for accelerated invoice factoring on one shipping lane may not be considered for accelerated invoice factoring on a different lane. Figure \ref{fig:decision-on-charges-for-factoring} captures the steps involved in the proposed approach. Smart contracts referred to here typically are part of the trade finance blockchain platform. Factors like charge type (planned or unplanned), computation type (fixed or variable), shipping lane, bank, carrier organization, and accuracy of the charge amount prediction impact this decision-making. Only a fraction of the predicted charge amount may be considered for factoring by the bank. Note that the carriers ideally want to include all the charges that are part of the invoice for accelerated invoice factoring. However, banks may not consider the charges like carrier-specific / lane-specific surcharges for factoring. The banks and carriers come together to decide the charges they want to include for factoring, the fraction of the charge amount considered for factoring, invoice discount/factoring fee, etc. and record the same on the trade finance blockchain platform. Banks might also directly interact with the AI service provider to evaluate the accuracy of the estimates and reward the AI service provider too.

\subsection{Final settlement of actual invoice amount}
\label{sec:final-settlement-actual-invoice}

At the end of the goods delivery by the carrier, the invoice amount in actuals is computed on the trade logistics network and shared with the banks on the trade finance network. The factoring company computes the sum of the installment amounts it paid to the carrier. If that amount is less than the amount to be paid against the invoice amount actuals (after subtracting the discounted amount from the actual invoice amount), then the factoring company processes the payment of the balance amount to the carrier. On the other hand, if that amount is more than the amount to be paid against the invoice amount actuals (after subtracting the discounted amount from the actual invoice amount), then the factoring company adjusts the excess amount paid against the future invoices for the same carrier.

\section{Implementation Details}
\label{sec:implementation}


In this section, we describe the implementation details of our proposed accelerated invoice factoring method for payment processing by considering sample independent blockchain consortium networks, TradeLens(TL) \cite{TLIBM} and We.Trade(WT) \cite{wt} that interoperate \cite{BVGCACM22} with each other. The sample WT (SWT) network helps banks facilitate invoice financing using our {\em accelerated invoice factoring} method. For simplicity, we assume that the sample TL (STL) connects suppliers, carriers, and shippers, while the sample WT connects banks and the carriers. We make use of the blockchain interoperability protocol \cite{ABGHKNPRVMIDDLEWARE19} for trusted data transfer between these desperate blockchain networks STL and SWT. Interoperability is the ability of blockchain networks to communicate for the transfer or exchange of data or value with assurances of validity. In particular, we propose to fetch the predicted carrier e-invoices from the trade logistics network, STL, using the trusted data transfer protocol \cite{ABGHKNPRVMIDDLEWARE19} and load them into the trade finance network, SWT, before triggering the accelerated invoice factoring process.

We consider a shipping lane with a transshipment port between the source and the destination ports. Moreover, we only consider the ocean carrier leg of the journey and no land carriers during freight transport for simplicity.

Figure \ref{fig:ourpaymentmethod} illustrates the communication between STL and SWT, and we describe below the steps involved to realize our proposed invoice finance method.
\begin{figure}[ht]
    \centering
    \includegraphics[width=0.99\columnwidth]{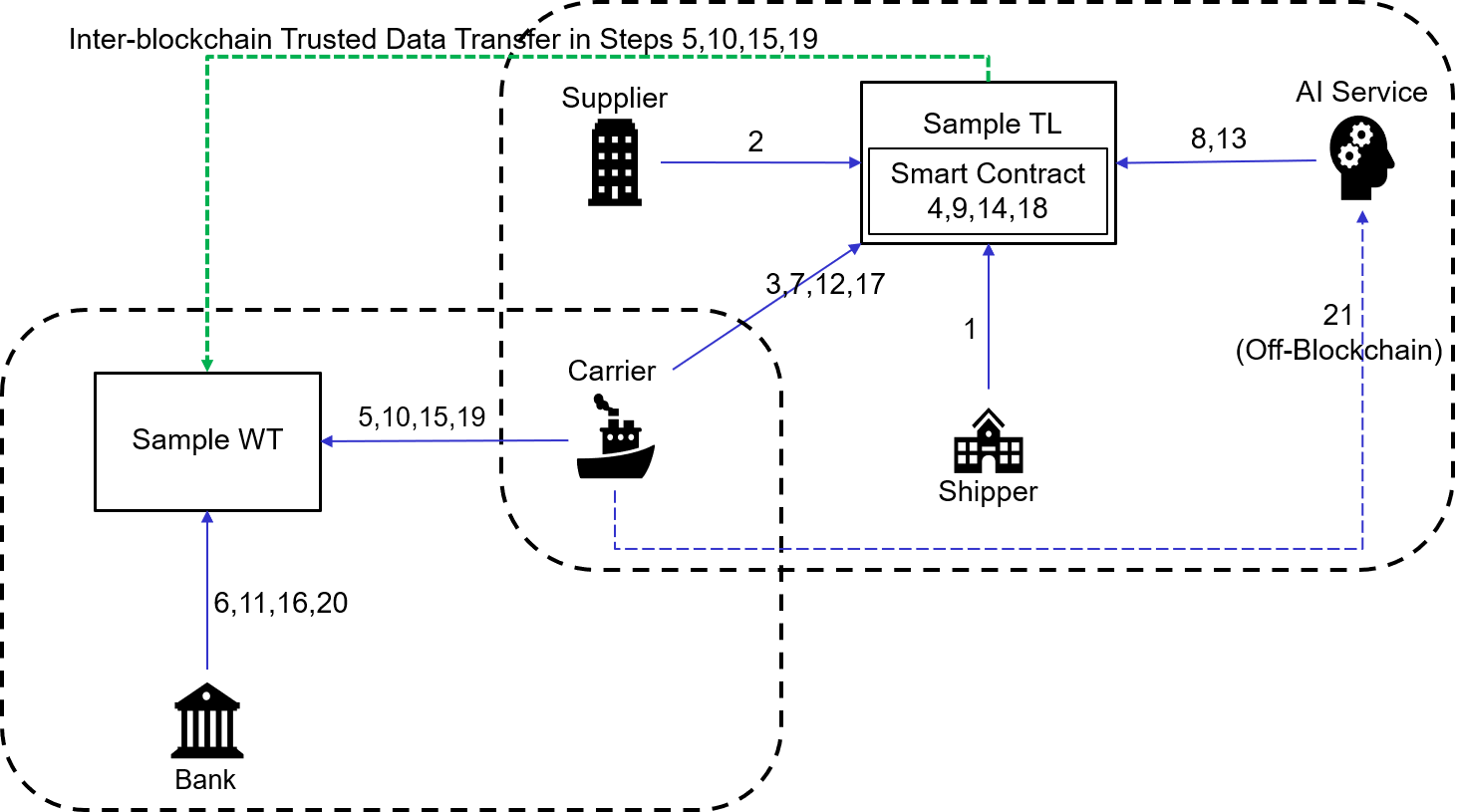}
    \caption{Steps in payment settlement using accelerated invoice factoring}
    \label{fig:ourpaymentmethod}
\end{figure}

\begin{enumerate}[leftmargin=*]
    \item Shipper initiates goods transfer by submitting a PO (purchase order).
    \item Supplier creates the consignment.
    \item Carrier emits milestone event {\em container loaded on Vessel}, after the bill of lading generation that conveys receipt of shipment.
    \item Smart Contract on STL generates a partial carrier e-invoice with actual values of planned charges.
    \item Carrier fetches partial e-invoice using the inter-blockchain trusted data transfer protocol from STL into SWT.
    \item Bank initiates the first chunk of invoice payment to the carrier organization.
    \item Carrier emits milestone event {\em departure from origin port} after dispatch of the consignment.
    \item AI service estimates and emits milestone event {\em  arrival at destination port}.
    \item Smart contract on STL updates partial carrier e-invoice to include actual values of planned charges and estimated values of unplanned charges.
    \item Carrier fetches updated partial e-invoice using the inter-blockchain trusted data transfer protocol from STL into SWT.
    \item Bank initiates one more chunk of invoice payment to the carrier organization for the estimated unplanned charges if it decides to pay additional invoice amount.
    \item Carrier emits milestone event {\em departure from transshipment port} at a transshipment location on the way to the destination port.
    \item AI service updates occurrence estimate of milestone event {\em arrival at destination port}.
    \item Smart contract on STL updates partial carrier e-invoice to include actual and estimated values of planned and unplanned charges.
    \item Carrier fetches updated partial e-invoice using the inter-blockchain trusted data transfer protocol from STL into SWT.
    \item Bank initiates one more chunk of invoice payment to carrier organization if the bank decides to pay more invoice amount.
    \item Carrier emits milestone event {\em arrival at destination port} and completes goods delivery to the shipper.
    \item Smart contract on STL generates final carrier e-invoice to include actual values of all the eligible planned and unplanned charges.
    \item Carrier fetches final e-invoice using the inter-blockchain trusted data transfer protocol from STL into SWT.
    \item Bank initiates final chunk of invoice payment to carrier organization if the sum of the chunks of invoice amounts paid until then is less than the discounted invoice amount.
    \item Carrier initiates reward for the AI service.
\end{enumerate}
The reward to the AI service can be off-blockchain processing. Similarly, the shipper pays the exact amount as in the e-invoice to the bank after $60$ days of goods delivery which can be another transaction on SWT with the shipper also being a network participant.
\section{AI model training}
\label{sec:modeltraining}

In this section, we discuss options for AI model training. The examples that we provide here are just for illustrative purposes. The actual model training depends on the structure of the charges component of the specific carrier invoice and the type of shipping milestone events used to compute those charges. We implemented the AI models in Python (version $3$) and using IBM Watson Studio AutoAI service \cite{WatsonAutoAI}.

We illustrate implementing the AI model by considering anonymized shipment details obtained from a trade logistics blockchain platform for few shipping lanes for two months, with a transshipment port between the export and import ports on a subset of these lanes. Hence, there will be only one leg of journey if there is no transshipment port between the origin and destination port or there will be two legs of journey with the first leg from the origin port to the transshipment port and the second leg from the transshipment port to the destination port. We consider the  {\em Dwell Excess Fee} \cite{DwellFee1, DwellFee2, DwellFee3} as one of the charge component of the carrier invoice for these shipping lanes. {\em Dwell time} is the time elapsed from the time a container arrives in a port to the time the container leaves the port (i.e., container waiting time at a given port). Let us say, there is charge of  $\$100$ per container on the first day past the set dwelling limit for a port terminal. Further, let us say that the charges increases by $\$100$ for each passing day (i.e., $\$200$ for the day $2$, $\$300$ for the day $3$, and so on). So, if the terminal dwelling limit is $1$ day for ocean carriers and the actual dwell time for a container is $3$ days, then the carrier invoice charge {\em Dwell Excess Fee} would be $\$300$ ($\$100$ for the day $2$ + $\$200$ for the day $3$).

For the shipping lanes we consider, we need to compute the {\em dwell time} for the transshipment ports that are between the origin and destination ports. Figure \ref{fig:event-types} captures the sample container tracking events during the freight transport from origin to the destination port. We predict the charge {\em dwell excess fee} first at the occurrence of the milestone event {\em Vessel Departure at Origin Port}. We again predict this same charge at the occurrence of the milestone event {\em Vessel Arrival at Transshipment Port}.


\begin{figure}[ht]
    \centering
    \includegraphics[width=0.99\columnwidth]{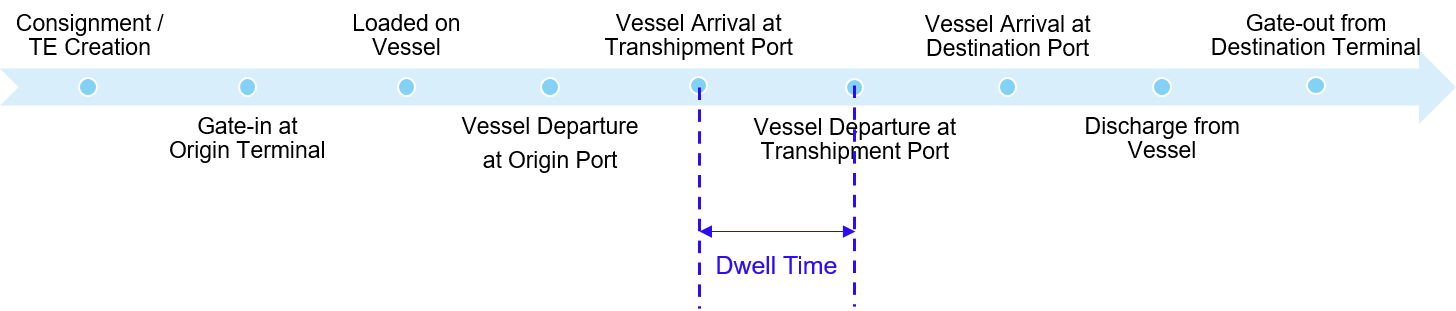}
    \caption{Sample container tracking milestone events}
    \label{fig:event-types}
\end{figure}


\subsection{Data characteristics}
\label{sec:data-characteristics}
We consider different types of data features for the model.
\begin{itemize}[leftmargin=*]
    \item Vessel characteristics: age, length, tonnage, capacity
    \item Port operations: number of berths, expected vessels, vessels in port
    \item Container milestone events (Planned \& Actual)
    \item Transport data: Scheduled transport plan of a consignment
\end{itemize}
The planned event data becomes available after the consignment creation. Later, the estimated events get emitted as the shipment progresses from the export to the import location. Finally, the actual event data is available at the occurrence of each milestone event. These events (planned or estimated or actuals) may get generated by different sources (e.g., carriers, drayage providers, port authorities, or origin-cargo-management teams). Moreover, any of these events may be emitted more than once by a given source on the occurrence of a milestone (e.g., drayage provider emits an estimated event and soon emits the same event again with an updated occurrence time), and we consider the latest reported event in that case.

One could consider external data like weather (geostrophic wind speeds, significant wave height, spectral peak wave period, vector mean wave direction), vessel position, and deviations/disruptions if such data is available. 

\subsection{RNN architecture for end-to-end prediction}
\label{sec:rnn-architecture}
\begin{figure}[ht]
    \centering
    \includegraphics[width=0.99\columnwidth]{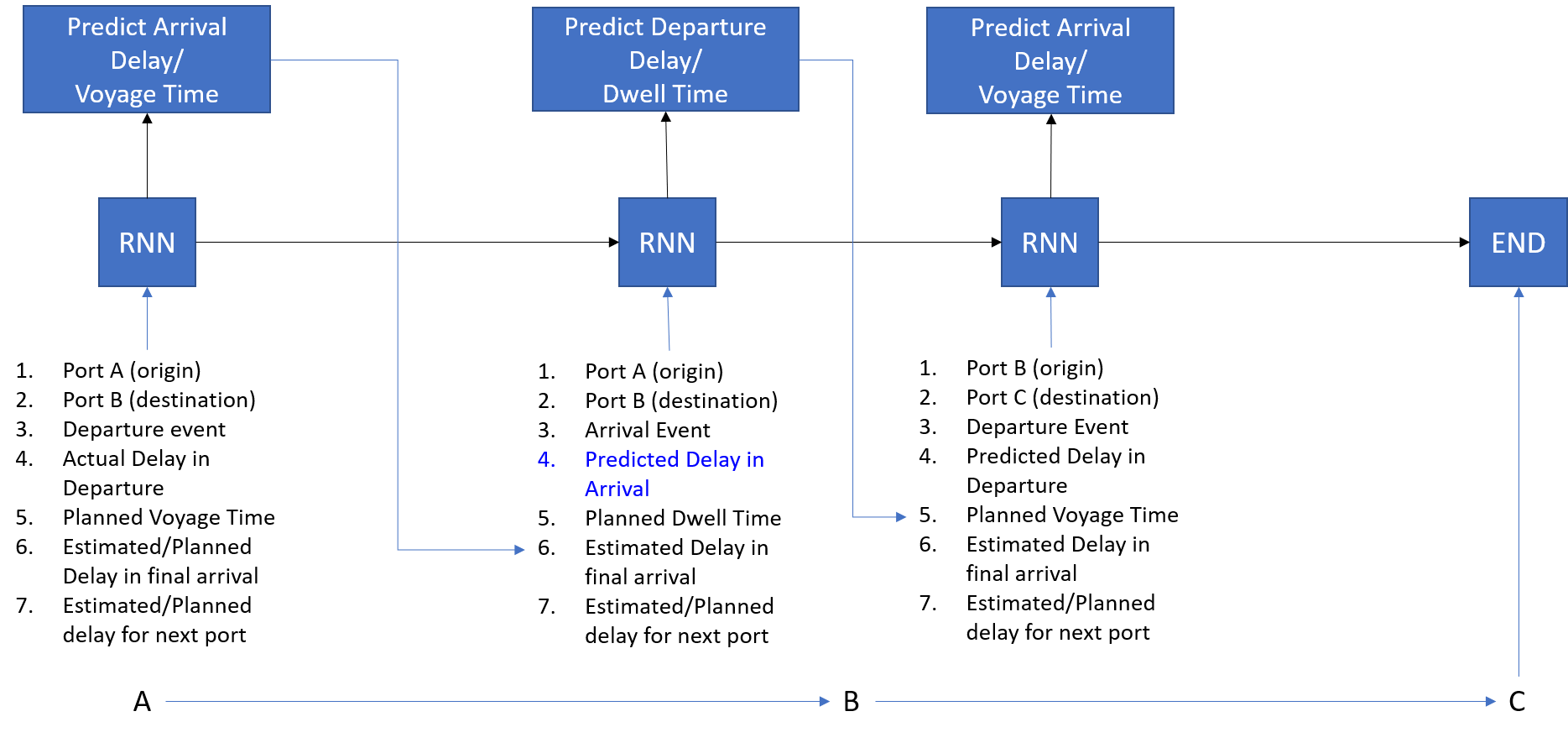}
    \caption{RNN classification model after departure from origin port}
    \label{fig:after-dept-at-origin}
\end{figure}
We use a recurrent neural network (RNN) \cite{EMNLP14, NIPS14} classifier to predict the dwell time. As the container moves from the origin port $A$  to the destination port $C$ via the transshipment port $B$, we carry out the following tasks: 
\begin{itemize}[leftmargin=*]
    \item {\em after departure from origin:} predict the delay in arrival at $B$, predict the voyage time (time from $A$ to $B$), predict the dwell time at $B$, and predict the delay in arrival time at $C$
    \item {\em after arrival at transshipment:} predict the delay in departure at $B$, predict the dwell time at $B$, and predict the delay in arrival time at $C$
    \item {\em after departure from transshipment:} predict the delay in arrival at $C$ and predict the voyage time (time from B to $C$)
\end{itemize}
Figure \ref{fig:after-dept-at-origin} captures the RNN classification model with a few sample features. And Figure \ref{fig:after-arrival-at-transshipment} captures the RNN classification model with similar set of features but the feature {\em Actual Delay in Arrival} replacing the feature {\em Predicted Delay in Arrival} in Figure \ref{fig:after-dept-at-origin}. The features that we have captured in the Figures \ref{fig:after-dept-at-origin} \& \ref{fig:after-arrival-at-transshipment} are only subset of the actual list of features that we have used and represent the important features used by the prediction model. Here is a brief description of these features:
\begin{itemize}[leftmargin=*]
    \item {\em Origin port:} vector representation of the origin port for the current leg of journey (e.g., origin port is A if the leg of journey is from port A to port B)
    \item {\em Destination port:} vector representation of the destination port (e.g., destination port is B if the leg of journey is from port A to port B) 
    \item {\em Event type:} either $0$ representing {\em Departure event} or $1$ representing {\em Arrival event}.
    \item {\em Planned voyage time / Dwell time:} it will be the {\em Planned voyage time} for the {\em departure events} and {\em dwell time} for the {\em arrival events}.
    \item {\em Estimated delay in final arrival:} Difference between the planned and estimated final arrival times.
    \item {\em Estimated delay for next port:} The difference between the planned and estimated arrival times at the destination port of current leg of journey.
\end{itemize}
Our RNN model classifies the predicted dwell times into class-1 or class-0 as below:
\begin{itemize}[leftmargin=*]
    \item class-1: if {\em dwell time} $>$ 24 hours
    \item class-0: otherwise
\end{itemize}

\begin{figure}[ht]
    \centering
    \includegraphics[width=0.99\columnwidth]{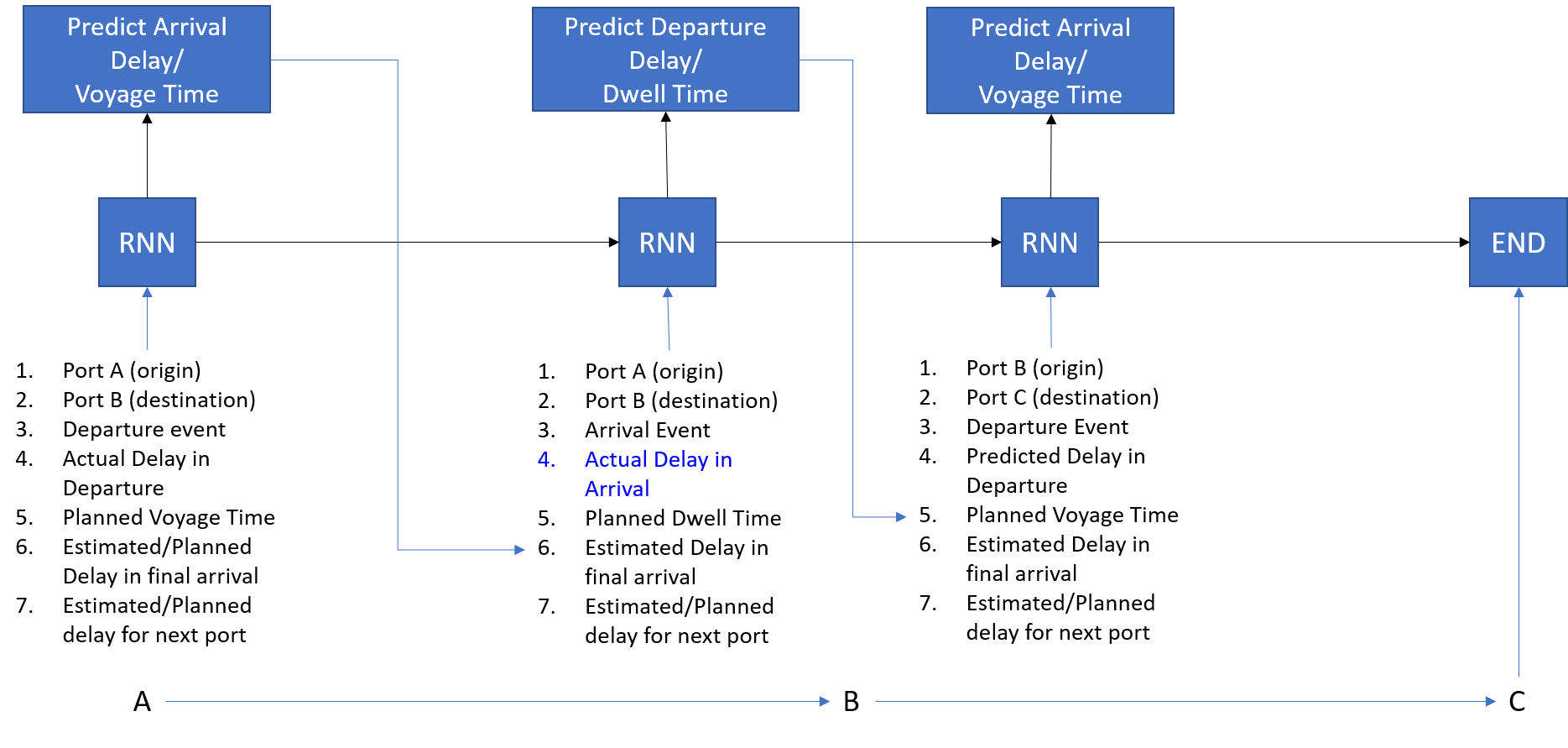}
    \caption{RNN classification model after arrival at transshipment port}
    \label{fig:after-arrival-at-transshipment}
\end{figure}
Though we have predicted the different tasks listed above, we only present the results for the {\em dwell time} prediction task in Table \ref{tab:result-one-transshipment-port}. Please note that we report the balanced accuracy, as there is class imbalance in the dataset. The balanced accuracy \cite{BalancedAccuracy} in binary classification problems is used to deal with imbalanced datasets. It is defined as the average of true positive rate and true negative rate. 

\begin{table}[h]
\caption{Results for Dwell Time Prediction with atmost 1 transshipment port}
\label{tab:result-one-transshipment-port}
\begin{tabular}{c|c|c|c}
\specialrule{.2em}{.1em}{.1em} 
\textbf{\begin{tabular}[c]{@{}c@{}}Model\\ Definition\end{tabular}}                   & \textbf{\begin{tabular}[c]{@{}c@{}}Prediction on \\ Milestone Event\end{tabular}} & \textbf{Model Type} & \textbf{\begin{tabular}[c]{@{}c@{}}Balanced \\ Accuracy (\%)\end{tabular}} \\ \specialrule{.2em}{.1em}{.1em} 
\begin{tabular}[c]{@{}c@{}}Binary\\ Classification\\ ( \textless 24 hrs)\end{tabular} & \begin{tabular}[c]{@{}c@{}}Departure from\\ Origin Port\end{tabular}              & Dwell Time          & 70.27                                                                     \\ \hline
\begin{tabular}[c]{@{}c@{}}Binary\\ Classification\\ ( \textless 24 hrs)\end{tabular} & \begin{tabular}[c]{@{}c@{}}Arrival at\\ Transshipment Port\end{tabular}           & Dwell Time          & 70.53                                                                     \\ \specialrule{.2em}{.1em}{.1em} 
\end{tabular}
\end{table}

Please note that it is easy to train a different model to classify if the dwell time is $>$48 hours or $>$72 hours. Using these models, one can predict if the dwell time for a given shipment is less than $1$ day, or $2$ days, or $3$ days, or more than $3$ days. And accordingly, the expected {\em Dwell Excess Fee} can be computed for that shipment using the predicted dwell time.

\subsection{Decision making by factoring organization}
\label{sec:factoring-org-decision}

By referring to the prediction accuracy in Table \ref{tab:result-one-transshipment-port}, the factoring organization can decide if a predicted charge amount can be considered for accelerated factoring at a given shipping milestone or wait for the occurrence of a subsequent milestone and decide based on the updated predicted charge value at that milestone. Prediction accuracy can be imported into the trade finance network from the trade logistics network as part of the trusted data transfer along with the predicted carrier invoice in Steps $5, 10, 15$ and $19$ of Section \ref{sec:implementation}. For instance, consider that our RNN model predicts the {\em dwell time} of a shipment to be more than 24-hours (and to be less than 48-hours using another RNN model). In other words, the predicted value of dwell time is $2$ days. Then, the smart contracts on the trade logistics blockchain generate the partial carrier invoice that includes the  {\em dwell excess fee} with predicted charge amount $\$100$.  As the accuracy of our RNN model is almost the same at both the milestone events, {\em departure at origin} and {\em arrival at transshipment} (refer to Table \ref{tab:result-one-transshipment-port}), the factoring organization could decide to release a chunk of the invoice amount corresponding to the predicted {\em dwell excess fee} at the occurrence of the milestone event {\em departure at origin} itself. It helps the carrier organization have access to the  {\em dwell excess fee} component even before the occurrence of {\em arrival at transshipment port} milestone event.

\subsection{Prediction with Multiple Transshipment Ports}
The same RNN model that we have presented in Figure \ref{fig:after-arrival-at-transshipment} can also be used to predict the dwell time task at multiple transshipment ports, if there are more than one transshipment port on the shipping lane from origin port to the destination port. We present the results for dwell time prediction task in Table \ref{tab:result-two-transshipment-port} for those lanes with two transshipment ports.

\begin{table}[h]
\caption{RNN classification model: Results for Dwell Time Prediction with 2 transshipment ports}
\label{tab:result-two-transshipment-port}
\begin{tabular}{c|c|c|c}
\specialrule{.2em}{.1em}{.1em} 
\textbf{\begin{tabular}[c]{@{}c@{}}Model\\ Definition\end{tabular}}                   & \textbf{\begin{tabular}[c]{@{}c@{}}Prediction on \\ Milestone Event\end{tabular}} & \textbf{Model Type} & \textbf{\begin{tabular}[c]{@{}c@{}}Balanced \\ Accuracy (\%)\end{tabular}} \\ \specialrule{.2em}{.1em}{.1em} 
\begin{tabular}[c]{@{}c@{}}Binary\\ Classification\\ ( \textless 24 hrs)\end{tabular} & \begin{tabular}[c]{@{}c@{}}Departure from\\ Origin Port\end{tabular}              & Dwell Time          & 75.19                                                                     \\ \hline
\begin{tabular}[c]{@{}c@{}}Binary\\ Classification\\ ( \textless 24 hrs)\end{tabular} & \begin{tabular}[c]{@{}c@{}}Arrival at 1st \\ Transshipment Port\end{tabular}      & Dwell Time          & 75.69                                                                     \\ \hline
\begin{tabular}[c]{@{}c@{}}Binary\\ Classification\\ ( \textless 24 hrs)\end{tabular} & \begin{tabular}[c]{@{}c@{}}Arrival at 2nd \\ Transshipment Port\end{tabular}      & Dwell Time          & 77.10                                                                     \\ \specialrule{.2em}{.1em}{.1em} 
\end{tabular}
\end{table}

As we can see in the table \ref{tab:result-two-transshipment-port}, that the Dwell Time prediction is improving as the freight transportation progresses from origin port to the destination port for these lanes.

\subsection{Miscellaneous recommendations}
\label{sec:recommendations}
Instead of using single RNN model for all shipping lane, one may want to train an independent RNN model for each shipping lane.  Additionally, one would be required to train different models for each milestone event corresponding to one or more charges of a carrier invoice.  Also, all the AI models need to be re-trained periodically so that it can capture the changes in operating conditions and external factors. In this paper, we have looked at classification models for dwell time prediction as dwell excess fee computation only required whether dwell time will be above certain threshold (24 hours) or not. If exact dwell time value is required for invoice computation then one can train a regression model instead of classification model.

\section{Conclusion}
\label{sec:conclusion}
We have provided the details of a blockchain-based trade logistics system for global trade that computes predicted carrier invoices in real-time as the freight transport progresses from origin to destination. Predicted carrier invoice at different shipping milestone events represents the accrued freight charges till that time plus the freight charges based on the predicted freight transport milestone events. The milestone event prediction service can be a third-party service external to the trade logistics blockchain network or part of it. Carrier organization presents the predicted carrier invoice to the factoring organization (bank or financial institution) to trigger invoice finance processing before the goods delivery completion. Factoring organizations may choose to release a chunk of the invoice amount at each freight transport milestone event based on the predicted carrier invoice as the freight transportation progresses from the origin to the destination. Thus, the carrier may access the invoice amount before the shipment of the goods is completed. It also benefits the factoring organization to earn more by higher invoice discounts. The accelerated invoice factoring method introduced also applies to land carriers (truck and rail) and air carriers in addition to ocean carriers.

\bibliographystyle{IEEEtran}
\bibliography{icbc}

\begin{thebibliography}{10}
\providecommand{\url}[1]{#1}
\csname url@samestyle\endcsname
\providecommand{\newblock}{\relax}
\providecommand{\bibinfo}[2]{#2}
\providecommand{\BIBentrySTDinterwordspacing}{\spaceskip=0pt\relax}
\providecommand{\BIBentryALTinterwordstretchfactor}{4}
\providecommand{\BIBentryALTinterwordspacing}{\spaceskip=\fontdimen2\font plus
\BIBentryALTinterwordstretchfactor\fontdimen3\font minus
  \fontdimen4\font\relax}
\providecommand{\BIBforeignlanguage}[2]{{%
\expandafter\ifx\csname l@#1\endcsname\relax
\typeout{** WARNING: IEEEtran.bst: No hyphenation pattern has been}%
\typeout{** loaded for the language `#1'. Using the pattern for}%
\typeout{** the default language instead.}%
\else
\language=\csname l@#1\endcsname
\fi
#2}}
\providecommand{\BIBdecl}{\relax}
\BIBdecl

\bibitem{NGSSSICBC21}
K.~Narayanam, S.~Goel, A.~Singh, Y.~Shrinivasan, and P.~Selvam, ``Blockchain
  based accounts payable platform for goods trade,'' in \emph{{IEEE}
  International Conference on Blockchain and Cryptocurrency, {(ICBC)}}, 2021.

\bibitem{factoring}
\BIBentryALTinterwordspacing
``Invoice factoring,'' {A}ccessed 10-Dec-2021. [Online]. Available:
  \url{https://en.wikipedia.org/wiki/Factoring_(finance)}
\BIBentrySTDinterwordspacing

\bibitem{letterOfCredit}
\BIBentryALTinterwordspacing
``U{N} {T}rade {F}acilitation {I}mplementation {G}uide. {L}etter of credit.''
  2019, {A}ccessed 10-Dec-2021. [Online]. Available:
  \url{https://tfig.unece.org/contents/letters-of-credit.htm}
\BIBentrySTDinterwordspacing

\bibitem{openaccount}
\BIBentryALTinterwordspacing
``U{N} {T}rade {F}acilitation {I}mplementation {G}uide. {O}pen {A}ccount.''
  2019, {A}ccessed 10-Dec-2021. [Online]. Available:
  \url{https://tfig.unece.org/contents/open-accounts.htm}
\BIBentrySTDinterwordspacing

\bibitem{TLIBM}
\BIBentryALTinterwordspacing
``{TradeLens: Digitizing The Global Supply Chain},'' {A}ccessed on 10-Dec-2021.
  [Online]. Available: \url{https://www.tradelens.com}
\BIBentrySTDinterwordspacing

\bibitem{NGSSCSCVBlockchain20}
K.~Narayanam, S.~Goel, A.~Singh, Y.~Shrinivasan, S.~Chakraborty, P.~Selvam,
  V.~Choudhary, and M.~Verma, ``Blockchain based e-invoicing platform for
  global trade,'' in \emph{{IEEE} International Conference on Blockchain
  {(Blockchain'20)}}, 2020.

\bibitem{bcforap}
\BIBentryALTinterwordspacing
``{\em Blockchain for Accounts Payable},'' {A}ccessed 10-Dec-2021. [Online].
  Available: \url{https://cporising.com/2017/05/
  25/blockchain-for-accounts-payable-an-introduction/}
\BIBentrySTDinterwordspacing

\bibitem{bcinar}
\BIBentryALTinterwordspacing
``{\em Blockchain in Accounts Receivable},'' {A}ccessed on 10-Dec-2021.
  [Online]. Available: \url{https://netsend.com/blog/
  blockchain-accounts-receivable/}
\BIBentrySTDinterwordspacing

\bibitem{wt}
\BIBentryALTinterwordspacing
``{We.Trade},'' 2019, {A}ccessed on 10-Dec-2021. [Online]. Available:
  \url{https://we-trade.com/}
\BIBentrySTDinterwordspacing

\bibitem{marcopolo}
\BIBentryALTinterwordspacing
``{Marco Polo Network},'' {A}ccessed 10-Dec-2021. [Online]. Available:
  \url{https://www.marcopolonetwork.com/}
\BIBentrySTDinterwordspacing

\bibitem{chang2020blockchain}
S.~E. Chang, H.~L. Luo, and Y.~Chen, ``Blockchain-enabled trade finance
  innovation: A potential paradigm shift on using letter of credit,''
  \emph{Sustainability}, vol.~12, no.~1, p. 188, 2020.

\bibitem{chiu2019blockchain}
J.~Chiu and T.~V. Koeppl, ``Blockchain-based settlement for asset trading,''
  \emph{The Review of Financial Studies}, vol.~32, no.~5, pp. 1716--1753, 2019.

\bibitem{bogucharskov2018adoption}
A.~Bogucharskov, I.~Pokamestov, K.~Adamova, and Z.~N. Tropina, ``Adoption of
  blockchain technology in trade finance process,'' \emph{Journal of Reviews on
  Global Economics}, vol.~7, pp. 510--515, 2018.

\bibitem{DemurrageCharge}
\BIBentryALTinterwordspacing
``Ocean carrier invoice: {D}emurrage charge,'' {A}ccessed 10-Dec-2021.
  [Online]. Available:
  \url{https://www.maersk.com/local-information/europe/norway/import}
\BIBentrySTDinterwordspacing

\bibitem{oceanfreightfactoring1}
\BIBentryALTinterwordspacing
``{See freight transportation factoring},'' {A}ccessed 10-Dec-2021. [Online].
  Available:
  \url{https://www.1sttruckingfactoring.com/sea_freight_factoring.htm}
\BIBentrySTDinterwordspacing

\bibitem{oceanfreightfactoring2}
\BIBentryALTinterwordspacing
``Freight factoring for marine transportation and ocean frieght cargo
  carriers,'' {A}ccessed 10-Dec-2021. [Online]. Available:
  \url{https://cfgbusiness.net/marinefactoring-oceanfunding.htm}
\BIBentrySTDinterwordspacing

\bibitem{oceanfreightfactoring3}
\BIBentryALTinterwordspacing
``Freight {F}actoring {F}or {T}ransportation {C}ompanies,'' {A}ccessed
  10-Dec-2021. [Online]. Available:
  \url{http://www.fnyo.com/transportation-factoring-company.htm}
\BIBentrySTDinterwordspacing

\bibitem{BVGCACM22}
\BIBentryALTinterwordspacing
R.~Belchior, A.~Vasconcelos, S.~Guerreiro, and M.~Correia, ``A survey on
  blockchain interoperability: Past, present, and future trends,'' \emph{{ACM}
  Comput. Surv.}, vol.~54, no.~8, pp. 168:1--168:41, 2022. [Online]. Available:
  \url{https://doi.org/10.1145/3471140}
\BIBentrySTDinterwordspacing

\bibitem{ABGHKNPRVMIDDLEWARE19}
E.~Abebe, D.~Behl, C.~Govindarajan, Y.~Hu, D.~Karunamoorthy, P.~Novotn{\'{y}},
  V.~Pandit, V.~Ramakrishna, and C.~Vecchiola, ``Enabling enterprise blockchain
  interoperability with trusted data transfer (industry track),'' in
  \emph{Proceedings of the 20th International Middleware Conference Industrial
  Track}.\hskip 1em plus 0.5em minus 0.4em\relax {ACM}, 2019, pp. 29--35.

\bibitem{WatsonAutoAI}
\BIBentryALTinterwordspacing
``Ibm watson studio autoai,'' {A}ccessed 10-Dec-2021. [Online]. Available:
  \url{https://www.ibm.com/in-en/cloud/watson-studio/autoai}
\BIBentrySTDinterwordspacing

\bibitem{DwellFee1}
\BIBentryALTinterwordspacing
``Container {D}well {F}ee set to begin {|} {P}egasus,'' {A}ccessed 10-Dec-2021.
  [Online]. Available:
  \url{https://www.pegasusmaritime.com/news/container-dwell-fees-nov-2021}
\BIBentrySTDinterwordspacing

\bibitem{DwellFee2}
\BIBentryALTinterwordspacing
``Ocean carrier charge: Container {D}well {F}ee,'' {A}ccessed 10-Dec-2021.
  [Online]. Available:
  \url{https://www.maersk.com/news/articles/2021/10/28/psw-emergency-fee-surcharge}
\BIBentrySTDinterwordspacing

\bibitem{DwellFee3}
\BIBentryALTinterwordspacing
``Port of {L}os {A}ngeles and {L}ong {B}each {D}well {T}ime {F}ees,''
  {A}ccessed 10-Dec-2021. [Online]. Available:
  \url{https://gcaptain.com/port-of-los-angeles-and-long-beach-dwell-time-fees-start-flying-today/}
\BIBentrySTDinterwordspacing

\bibitem{EMNLP14}
K.~Cho, B.~van Merrienboer, {\c{C}}.~G{\"{u}}l{\c{c}}ehre, D.~Bahdanau,
  F.~Bougares, H.~Schwenk, and Y.~Bengio, ``Learning phrase representations
  using {RNN} encoder-decoder for statistical machine translation,'' in
  \emph{Proceedings of the 2014 Conference on Empirical Methods in Natural
  Language Processing, {EMNLP} 2014, October 25-29, 2014, Doha, Qatar, {A}
  meeting of SIGDAT, a Special Interest Group of the {ACL}}.\hskip 1em plus
  0.5em minus 0.4em\relax {ACL}, 2014, pp. 1724--1734.

\bibitem{NIPS14}
\BIBentryALTinterwordspacing
I.~Sutskever, O.~Vinyals, and Q.~V. Le, ``Sequence to sequence learning with
  neural networks,'' in \emph{Advances in Neural Information Processing Systems
  27: Annual Conference on Neural Information Processing Systems 2014, December
  8-13 2014, Montreal, Quebec, Canada}, 2014, pp. 3104--3112. [Online].
  Available:
  \url{https://proceedings.neurips.cc/paper/2014/hash/a14ac55a4f27472c5d894ec1c3c743d2-Abstract.html}
\BIBentrySTDinterwordspacing

\bibitem{BalancedAccuracy}
\BIBentryALTinterwordspacing
``What is balanced accuracy?'' {A}ccessed 10-Dec-2021. [Online]. Available:
  \url{https://statisticaloddsandends.wordpress.com/2020/01/23/what-is-balanced-accuracy/}
\BIBentrySTDinterwordspacing

\end{thebibliography}

\end{document}